# Cold Regions Science and Technology

# Experimental investigations on the characteristics of snow accretion using EMU-320 model train

--Manuscript Draft--

| | |
|---|---|
| Abstract: | The driving condition on railways is negatively influenced by harsh weather such as snow because snow accumulation on the main parts of rolling stock leads to train malfunction and degrades driving performance. This paper presents a snow accretion test conducted in a climate wind tunnel to investigate the icing process on a model train. The model used within this experiment was the cleaned-up and 2/3-scaled version of EMU-320, which is a high-speed train in Korea. The model was designed without an electronic power source or heat source so that the wheels did not rotate and snow accretion on the model did not occur due to heat sources. To investigate snow accretion, four cases with different ambient temperatures were considered in the climate wind tunnel on Rail Tec Arsenal. Before analyzing the snow accretion on the train, the snow flux and liquid water content of snow were measured so that they could be used as the input conditions for the simulation and to ensure the analysis of the icing process was based on the characteristics of the snow. Both qualitative and quantitative data were obtained, whereby photographs was used for qualitative analysis, and the density of the snow sample and the thickness of snow accreted on the model were used for quantitative analysis. Based on the visual observations, it was deduced that as the ambient temperature increased, the range of the snow accreted was broader. The thickness of snow accreted on the model nose was the largest on the upper and lower part at -3 o C, and on the middle part at -5 o C. Additionally, the cross section of snow accreted was observed to be trench-like. Similar icing processes were observed to occur on the slope of nose. Snow accreted on all components of the bogie, and for all cases, the thickness of snow at wheel was the largest at an arc angle of 40 to 70 o . These detailed data of experimental conditions can be applied as an input to simulations to improve simulations of ice conditions. Thus, they can facilitate the development of appropriate anti-icing designs for trains |
| | Prof. Yu is a researcher conducting a study on snow drifting on roof. Prof. Yu is now a Head of Department of architectural engineering, Southwest Jiaotong University, China that has been conducted research on snow drifting. Also, Prof. Yu has a good understanding of the characteristics of the snow through numerous experiments and simulations. |
| | László E Kollár, Ph.D<br>Professor, Eötvös Loránd University<br>laszlo_kollar@uqac.ca<br>Prof. Kollár is a researcher on snow and icing accretion. Prof. Kollár has a lot of experience with snow accretion test. Author suggests Prof. Kollár as a review who can review test process and data obtaining method. |
| | Gisella Tomasini, Ph.D<br>Professor, Polytechnic of Milan<br>gisella.tomasini@polimi.it<br>Prof. Tomasini is one of the researchers with a lot of experience in experimental research. Prof. Tomasini has been conducted full-scale wind tunnel test and show and published many researches. Author suggests Prof. Tomasini as a review who can review test process and data obtaining method. |
| | Hisato Matsumiya, Ph.D<br>Researcher, Central Research Institute of Electric Power Industry<br>hisato-m@criepi.denken.or.jp<br>Mr. Matsumiya is a research science in Central Research Institute of Electric Power Industry, Tokyo, Japan. Mr. Matsumiya has been conducted researches about ice accretion. Author suggests him as a reviewer who can review snow accumulation process. |

July 13, 2022

Jukka Tuhkuri
Editor-in-Chief
*Cold Regions Science and Technology*

Dear Editor:

I wish to submit a research paper for publication in *Cold Regions Science and Technology,* titled "An experimental study on the characteristics of snow accretion using an EMU-320 model train." The paper was coauthored by Wan Gu Ji, Soonho Shon, Song Hyun Seo, and Beomsu Kim.

This study presents a snow accretion test on a cleaned-up and 2/3-scaled version of the EMU-320, the high-speed train in Korea. Our experiment was conducted in a climate wind tunnel to investigate the icing process on the model of the train. We believe that our study makes a significant contribution to the literature because trains are one of the most widely used transportation methods worldwide, although extreme weather events such as snow and ice have a negative impact on rolling stock such as trains. The driving condition on railways is negatively influence by harsh weather such as snow, because snow accumulation on the main parts of rolling stock leads to train malfunction and degrades driving performance. Therefore, we investigated the snow accretion and icing processes of a train model, considering four cases with varying ambient temperatures, by conducting experiments in the climate wind tunnel of Rail Tec Arsenal in Austria. Within this icing wind tunnel, a low temperature of the wind tunnel can be maintained and snow can be sprayed using a snow nozzle and ice rig. We designed our model train without an electronic power source or heat source to avoid the wheels rotating or snow collecting on the model due to heat sources. We obtained both qualitative and quantitative data, i.e., 3D scan data and photographs, and the density of the snow sample and thickness of the snow accreted on the model, respectively. We found that, as the ambient temperature increased, the range of thicknesses of snow accreted was broader. These detailed data of experimental conditions can be applied as an input to simulations to improve simulations of ice conditions. Thus, they can facilitate the development of appropriate anti-icing designs for trains

This manuscript has not been published or presented elsewhere in part or in entirety and is not under consideration by another journal. We have read and understood your journal's policies, and we believe that neither the manuscript nor the study violates any of these. Details about competing interests are provided separately.

Thank you for your consideration. I look forward to hearing from you.

Sincerely,
Kyuhong Kim
Institute of Advanced Aerospace Technology, Seoul National University
1, Gwanak-ro, Gwanak-gu, Seoul, 08826, Republic of Korea

aerocfd1@snu.ac.kr

# Experimental investigations on the characteristics of snow accretion using the EMU-320 model train

Wan Gu Ji[a], Soonho Shon[a], Song Hyun Seo[a], Beomsu Kim[b], Kyuhong Kim[a,c*]

[a] Department of Aerospace Engineering, Seoul National University, 1, Gwanak-ro, Gwanak-gu, Seoul, 08826, Republic of Korea

[b] Department of Railroad Vehicle System Engineering, Korea National University of Transportation, 156 Cheoldobang-mulgwan-ro, Uiwang-si, Gyeonggi-do, 16106, Republic of Korea

[c] Institute of Advanced Aerospace Technology, Seoul National University, 1, Gwanak-ro, Gwanak-gu, Seoul, 08826, Republic of Korea

[*] **Corresponding author.** Email address: aerocfd1@snu.ac.kr (Kyuhong Kim)

**Abstract**

The driving condition on railways is negatively influenced by harsh weather such as snow because snow accumulation on the main parts of rolling stock leads to train malfunction and degrades driving performance. This paper presents a snow accretion test conducted in a climate wind tunnel to investigate the icing process on a model train. The model used within this experiment was the cleaned-up and 2/3-scaled version of EMU-320, which is a high-speed train in Korea. The model was designed without an electronic power source or heat source so that the wheels did not rotate and snow accretion on the model did not occur due to heat sources. To investigate snow accretion, four cases with different ambient temperatures were considered in the climate wind tunnel on Rail Tec Arsenal. Before analyzing the snow accretion on the train, the snow flux and liquid water content of snow were measured so that they could be used as the input conditions for the simulation and to ensure the analysis of the icing process was based on the characteristics of the snow. Both qualitative and quantitative data were obtained, whereby photographs was used for qualitative analysis, and the density of the snow sample and the thickness of snow accreted on the model were used for quantitative analysis. Based on the visual observations, it was deduced that as the ambient temperature increased, the range of the snow accreted was broader. The thickness of snow accreted on the model nose was the largest on the upper and lower part at -3 ºC, and on the middle part at -5 ºC. Additionally, the cross section of snow accreted was observed to be trench-like. Similar icing processes were observed to occur on the slope of nose. Snow accreted on all components of the bogie, and for all cases, the thickness of snow at wheel was the largest at an arc angle of 40 to 70 °. These detailed data of experimental conditions can be applied as an input to simulations to improve simulations of ice conditions. Thus, they can facilitate the development of appropriate anti-icing designs for trains

**Keywords**: High-speed train, Snow accretion, Dry snow, Climate wind tunnel, Liquid water content, Snow distribution

## 1. Introduction

With the development of railway operation systems and design, the advantages of trains such as safety and punctuality have increased. This has resulted in trains being one of the most widely-used transportation methods worldwide. However, extreme weather events such as snow and ice have a negative impact on rolling stock. Many researches have reported that snowy weather badly affects not only the driving environment, such as the train signal system (Zaki (2019)) and rail icing (Kim (2015)), but also the performance of the train. There are various devices to ensure safe operation of trains, including brake calipers, dampers, and springs, yet these all generate heat continuously. The heat generated while driving melts the snow, and snow particles subsequently collide with the melted snow, causing ice agglomeration. Eventually, ice attaches to devices which can lead to difficulties in maintenance and a degradation of braking performance (Kloow (2011)). In addition, ice adversely affects the damping system and causes vibration of train body, which results in an uncomfortable ride (Giappino (2016)). Furthermore, quantitative analysis in four European countries demonstrated that extreme weather can delay train operation by up to 79% (Ludvigsen (2014)), and it was concluded that functional failure negatively affects normal train operation.

To solve this issue, a number of researches on snow accretion have been conducted over the past few decades, based on both simulation and experimentation.

Simulation studies include those of Ansari (2015) and Pankajakshan (2011), in which the movement characteristics of solid particles were simulated to analyze the two-phase flow field composed of solid-incompressible flow using the Euler–Lagrangian method. Particle collision models were implemented in a time-accurate flow solver, which had a good ability to capture the physics of particle movement. Moreover, the Discrete Phase Model (DPM) method has been applied to the Euler–Lagrangian method in order to predict snow concentration on the wheels, brakes, electromotor, and bogie frame (Xie (2017)). To verify de-icing performance, numerous simulation studies have been conducted. As one of them, the effect of de-icing has been studied using passive control methods, such as by changing the geometry of the deflector (Wang (2018)).

As simulation research can be done efficiently in a short time and with low computational cost, experimental studies of snow accretion on trains have significance by itself. Most importantly, results from tests can be used to investigate the cause of snow accretion on trains, highlighting the fact that snow accretion on trains is caused by various factors including complex flowfields and the characteristics of snow. Furthermore, although the cost for testing is considerable, results from experiments are useful for validation of simulations. This is because the experimental conditions including snow flux can be measured and applied as an input to numerical analysis to improve simulation results, and results from experiments and simulations can be directly compared.

Within experimental studies, various methods have been implemented. Zhou (2014) selected particles that can replace snow based on similarity parameters such as friction velocity, density ratio, and Froude number. Additionally, accretion tests were conducted using substituted particles and the results were compared with the accretion range predicted by simulations (Wang (2019)). Although these methods were successful in qualitative comparison of the simulation and experimentation, the substituted particles did not fully reflect the characteristics of snow, which can vary sensitively depending on the ambient temperature and humidity. Therefore, there are limitations associated with the method of using substituted particles, in that it is impossible to quantitatively predict the thickness of the snow layer.

Studies have been conducted outdoors using natural snow (Sakamoto (2000); Admirat (2008). However, most of the studies conducted outdoors have focused on power cable accretion analysis; moreover, outdoor measurements are limited by natural conditions that are difficult to reproduce. To overcome the limitation of outdoor experiments, climate wind tunnels have been used for research in the past few decades. It is reliable to use climate wind tunnels to investigate snow accretion because the experimental conditions can accurately reproduce outdoor environments. Furthermore, an experimental study conducted in a climate wind tunnel can obtain continuous, stable experimental results; moreover, the results can be reproduced.

Most previous studies using climate wind tunnel have been performed on a small scale,

such as snow accretion tests on airfoil (Shin (1992); Broeren (2004); Li (2014)) and evaporative cooling tests on miniature buildings (Zhang (2015)), as it is easier to obtain such experimental equipment and achieve operation compared to large-scale experiments. Recently, with the development of weather simulation technology, large-scale experimentation on train has been conducted. Bucek (2017) analyzed the effect of snow accretion on the bogie and windshield of a train during operation. However, this research focused on the influence on train operation caused by the malfunction of each device; the analysis of snow accretion was not examined directly. Wang (2022) proposed anti-icing designs on trains and discussed their effect based on comparative experiments. However, this was conducted to check the performance of the designed devices. Furthermore, insufficient research has been conducted to show the quantitative data from the experimentation.

Above all, previous studies have rarely described the characteristics of snow. It is meaningful to present the properties of snow, because the size of snow and how much water the snow particle contains determine ice agglomeration process. Furthermore, these obtained data by measuring can be used as numerical conditions to improve simulation results. Second, incoming snow flux must be measured before the experiment. Any experiment installation can not produce a uniform snow distribution, which is a typical limitation in climate wind tunnel. Therefore, the distribution of sprayed snow should be presented. Finally, the information of snow layer is barely presented. The snow layer is exposed to various environments such as heat transfer and collision of snow driven by flow. In this process, the characteristics of snow layer such as shape and density can be changed.

In this study, experimentation of snow accretion on a model train was conducted using a climate wind tunnel, in order to quantitatively obtain data regarding snow accretion on each location of the train, such as the nose and bogie, and to analyze the results according to the ambient temperature. Experiments were conducted in the icing wind tunnel of Rail Tec Arsenal (RTA) in Austria, where the wind tunnel can be maintained at a low temperature and snow can be sprayed using a snow nozzle and an ice rig. Before performing the experiment, the environment and snow spray conditions in the wind tunnel were measured. Snow was collected to measure the liquid water content. Finally, qualitative and quantitative results for snow accretion were obtained to identify the density and thickness of accreted snow.

## 2. Experimental setup
### 2.1 Test Facility

In order to simulate snowy weather conditions, a climate wind tunnel was used. The climate wind tunnel (CWT) in the RTA is a closed-loop wind tunnel that can control low and high temperatures and various environmental conditions, such as solar, rain, and snow. Additionally, the CWT in RTA has a quality management system certified to DIN ISO 9001 standard. The CWT is broadly used for icing tests, cold start testing of road vehicles and aircraft, and air conditioning tests in cockpits. The specifications for the CWT are given in Table 1. The CWT is operated by an interactive automatic control system that can set a certain temperature and wind speed. Particularly, the user can monitor all the data acquired by the CWT in real time in the preparation room.

Table 1 Climate wind tunnel specification

| Description | |
|---|---|
| Test section | |
| Width | 5.1m |
| Height | 6.0m |
| Length | 33.8m |
| Temperature range | -45°C~60°C |
| Maximum wind speed | |
| At -20 °C | 40m/s |
| At -30 °C | 30m/s |
| Turbulence intensity | 0.6% (at 10m/s) |
| | 1.2% (at 30m/ |

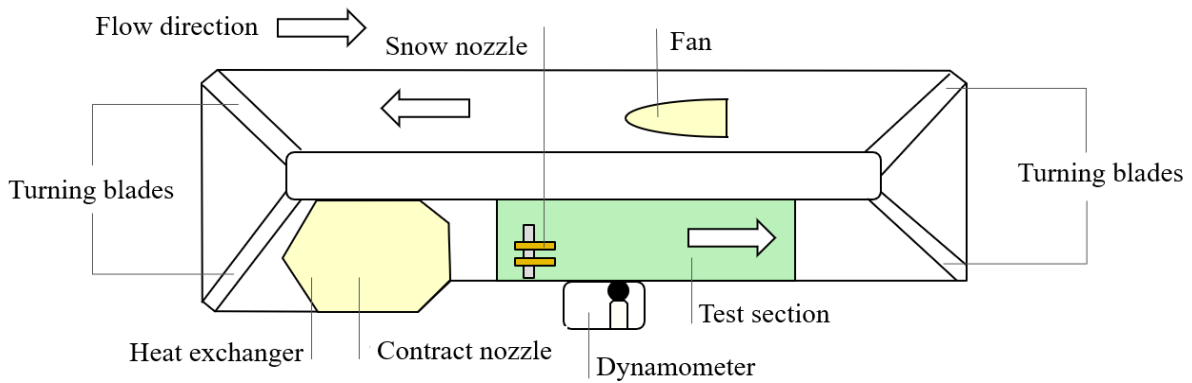

Figure 1 Schematics of climate wind tunnel in RTA

Before the snow accretion test, the temperature and speed should be stabilized. Therefore, the temperature and speed were measured during the experiment to obtain the environmental maintenance capacity of the wind tunnel. The environmental ambient temperature was -3 °C and the target wind speed was 30 m/s, i.e., the maximum wind speed of the CWT. Measurement was obtained every 10 s and the experiment was conducted for 30 min. The recorded temperatures and velocities are shown in Figure 2.

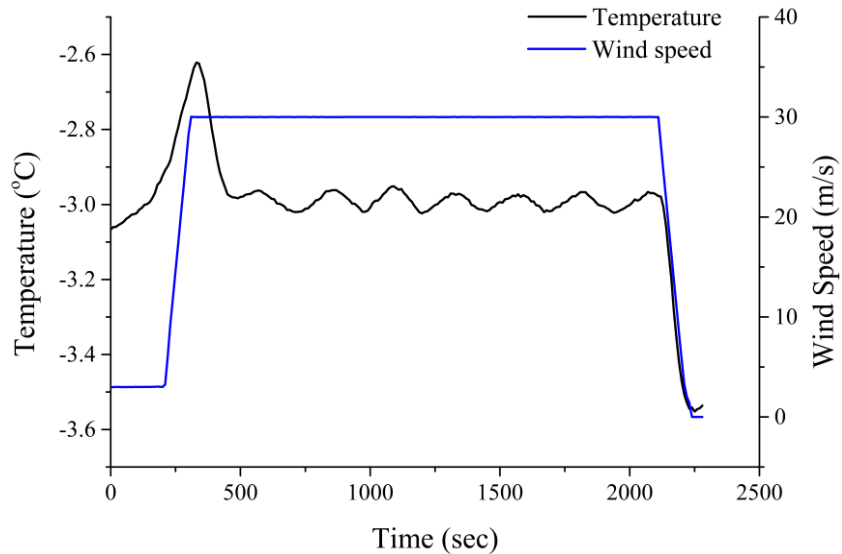

**Figure 2** Maintenance capacity of velocity and temperature in the climate wind tunnel

The experiment comprised the following variations in flow velocity: 210 s stationary, 90 s of acceleration, 1800 s of maintaining the flow velocity at 30 m/s, and 140 s of deceleration. Furthermore, it was found that the temperature increased by 0.4 ºC in the acceleration stage, but exhibited a small perturbation of ±0.05 ºC in the uniform velocity stage. Thus, through these measurements, it was confirmed that the CWT can produce a constant velocity and temperature during the experiment.

For spraying the snow, mobile snow nozzles were used. There were two types of pipes in the snow nozzle, one for supplying subcooled water and the other for supplying compressed air. In a previous study (Kozomara (2021)), the liquid water content (LWC) of snow particles and their median volume diameter (MVD) were determined by controlling the amount of water and the pressure of compressed air. Therefore, the flow rate of water and the pressure of compressed air can be set as input values, and the user can monitor the actual flow rate of water and pressure in real time. In this study, six nozzles were used to produce the snow particles. As shown in Figure 3, three nozzles were placed at both the top and bottom, and the distances between the nozzles were set to 800 mm. The lower nozzles were mounted 400 mm above the ground. This arrangement enabled spraying the entire model, including the bottom area of the bogie.

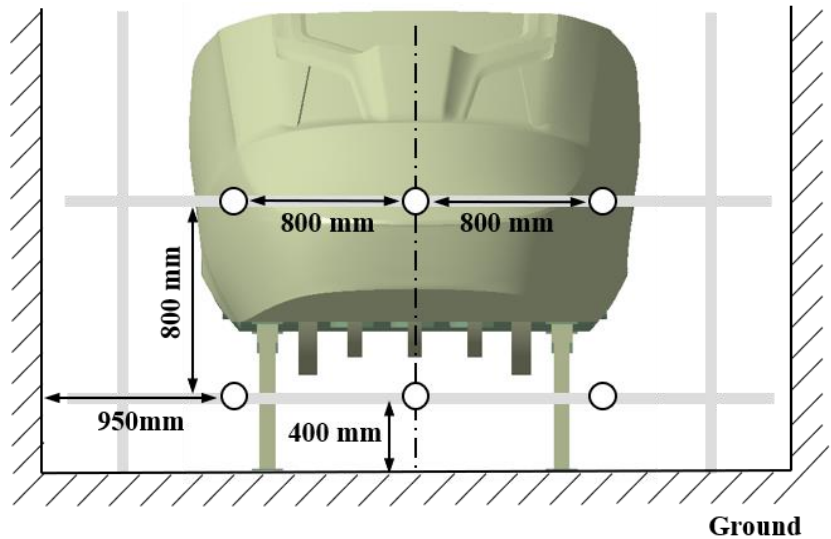

**Figure 3** Schematics of nozzle arrangement

## 2.2 Model Description

As shown in Figure 4, the testing model was a cleaned-up version of the EMU-320 high-speed train, scaled by 2/3. The length, width, and height of the testing model were 8.0 m, 2.0 m, and 2.2 m, respectively, and the model train was placed at a distance 4.4 m from the snow nozzle. As the experiment mainly focused on snow and ice accretion on the nose and bogie, the upper part of model was cut off in consideration of the blockage ratio and the manufacturing cost. The bogie is in the cavity area that airflow and snow particles are likely to enter.

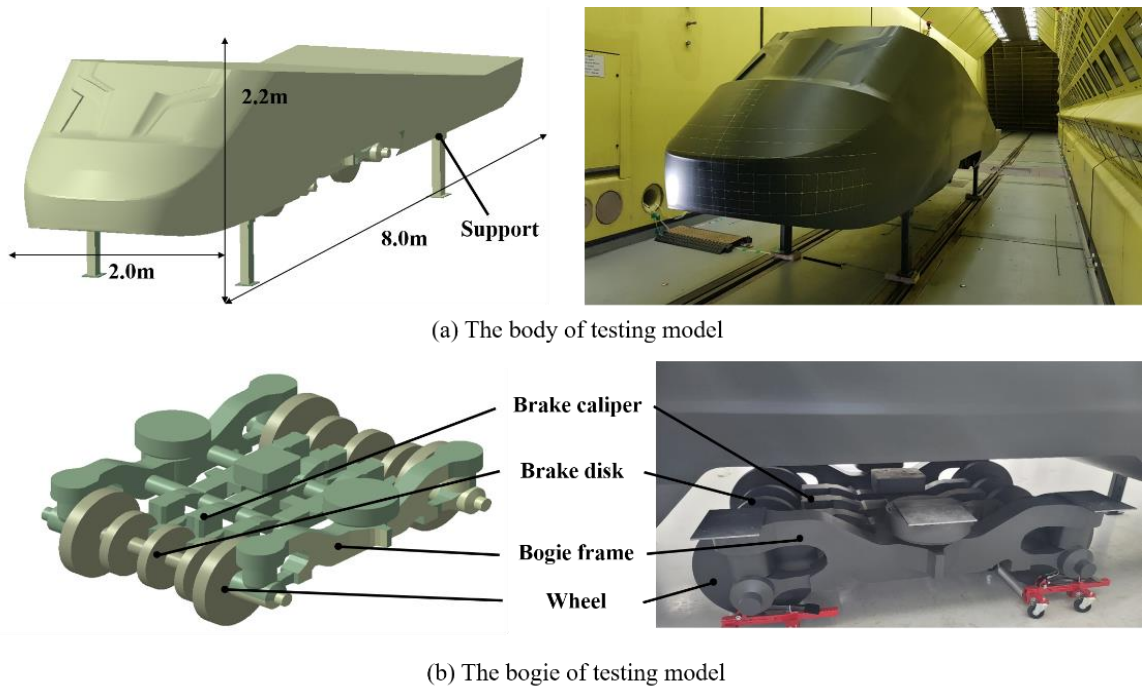

(a) The body of testing model

(b) The bogie of testing model

**Figure 4** Diagrams (left) and pictures (right) of the model

The bogie consists of wheels, brake calipers, brake discs, and the bogie frame, as shown in Figure 4 (b). This is simplified to easily investigate snow accretion. It is worth noting that the test model is non-powered so that wheels were not operational, and no heat sources were present in the bogie.

The support was made to control the height of model. In this study, the ground is a stationary wall, and previous studies have noted that ground effects, such as gradually thicker boundary layers on the ground and increased turbulence levels, can lead to high interference between the bogie and the stationary wall (Tyll (1996); Kwon (2001)). This setting is acknowledged within the limitations of the experimental setup. In order to weaken the influence of movement on the snow particles according to the stationary ground effect, the model was set to a height of 0.7 m above the ground.

Finally, the model was made of chemical wood and was painted black to easily check whether snow particles were attached or not. Particularly, a grid was drawn on the surface of body to easily obtain the data of the range of snow accretion. The gap between the grid was 10 cm in both the horizontal and vertical directions.

**2.3 Experimental Condition**

In Table 2, details of the experimental conditions are provided. Four cases were conducted with a wind speed of 30 m/s for 30 min. Since the snow particles containing a certain amount of moisture were inside the CWT during all cases, the relative humidity was maintained at approximately 85 %. For each case, the experimental condition had different ambient temperatures of -3, -5, -7, and -9 ºC.

Table 2 Operating conditions for experimental cases

| Exp. No | Temperature (ºC) | Relative humidity (%) | Flow velocity (m/s) | Running time (sec) | Water flow (L/h) | Pressure of compressed air (bar) |
|---|---|---|---|---|---|---|
| 1 | -9 | >85 | 30 | 1800 | 80 | 6.5 |
| 2 | -7 | >85 | 30 | 1800 | 80 | 6.5 |
| 3 | -5 | >85 | 30 | 1800 | 80 | 6.5 |
| 4 | -3 | >85 | 30 | 1800 | 80 | 6.5 |

For the acquisition of data, Exp. 1 and 4 were conducted twice. However, Exp. 2 and 3 could not be repeated because of the long experimental time and high operation costs; this may lead to a level of uncertainty. Nevertheless, the experiment was sufficient to explain the snow accretion on the train according to the ambient temperature, and the acquired data was expected to be used as a sample for numerical analysis to validate simulations.

The set values for spraying subcooled water were the same for all experimental cases. The water flow was set to 80 L/h and the pressure of the compressed air to 6.5 bar. Based on this spraying setting, the MVD of the particles was measured using the Spraytec system from Malven Instruments (Ferschitz (2017)). Figure 5 shows the MVD spectrum.

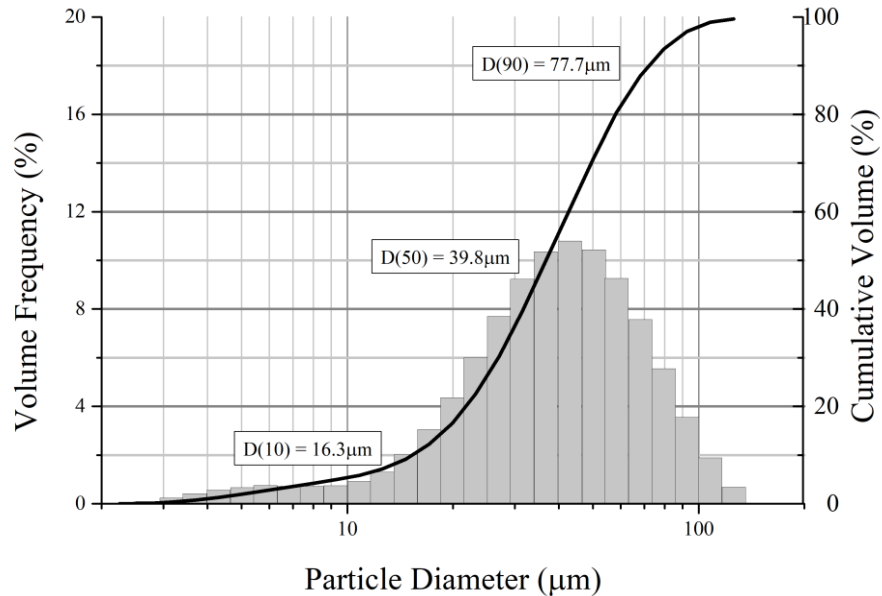

**Figure 5** Median volume diameter (MVD) spectrum for setting condition of 80 L/h of water flow and 6.5 bar of compressed air.

Figure 5 shows that 90 % of snow particles had a diameter smaller than 77.7 μm, and the most common particle size was a diameter between 40 and 50 μm, accounting for more than 20 % of total. Using this distribution, experiments were performed under the conditions presented in Table 2.

### 3. Measurement method
### 3.1 Flowfield

Flow characteristics on the bottom of the bogie were complex due to the geometry of the bogie and the boundary layer growing on the surface of model. Moreover, the snow particles were driven by flow and this can affect the particle concentration on bottom of the bogie. Thus, it is crucial to obtain the flow velocity on the bottom of the bogie. In this study, the flow velocity was measured at four locations, which are shown in Figure 6.

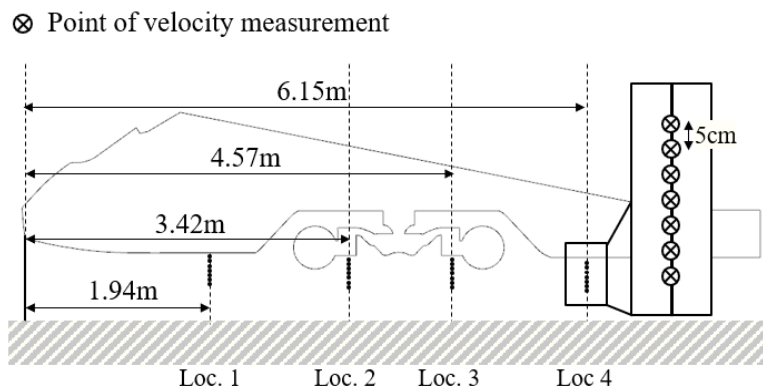

**Figure 6** Locations of velocity measurement points

To measure flow velocity, pitot tubes were used. A pitot tube is an intrusive pressure-sensitive instrument that can measure the flow velocity accurately when the flow direction and the measurement direction of the pitot tube are parallel. In this study, seven pitot tubes were installed at each location to measure the flow velocity on the bottom of model. The gaps between the pitot tubes were set to 5 cm to minimize the interference between the pitot tubes. The measuring range of the vertical length was 35 cm from the surface of model, and the installation scene is shown in Figure 7. To digitize the signal obtained from the pitot tubes, NetScanner 9216 was used, and calibration of the measuring velocity was conducted before the experiment using the open-looped wind tunnel. Since pitot tubes measure static pressure and dynamic pressure, 14 channels were used. The velocity was calculated from the Bernoulli equation using the averaged density data in the CWT.

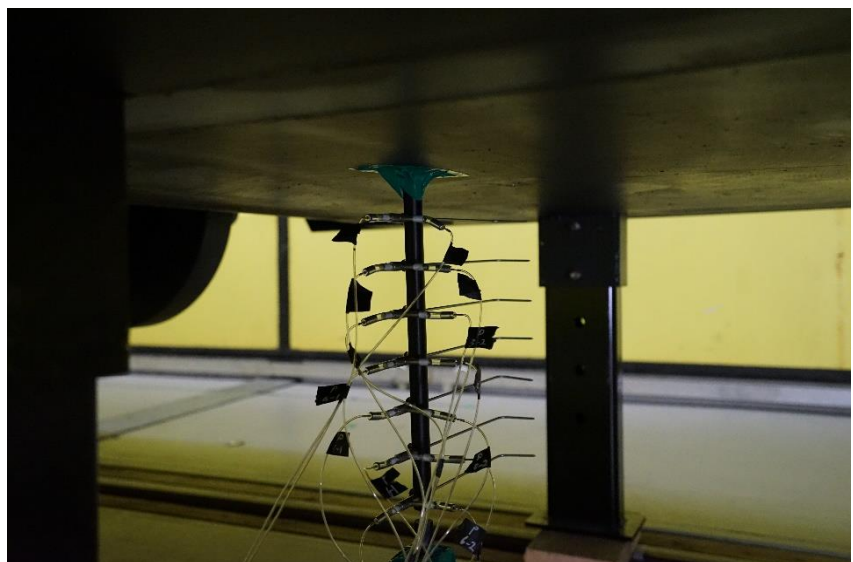

**Figure 7** Bundle of pitot tubes

Subsequently, nine stainless tubes were embedded in the model to measure the surface pressure of model. The length and width of the stainless tubes were 50 mm and 1.6 mm, respectively. Using a flexible rubber tube, each stainless tube was connected to a pressure scanner. The reference pressure was measured using a pressure gauge located on the CWT wall, 25 m away from the inlet. The locations of the pressure measurements are shown in Figure 8. Using the measured pressure at each location and the reference pressure, the velocity ratio $u$ and the coefficient of surface pressure $C_p$ was calculated as follows:

$$u = \frac{U}{U_{ref}}, C_p = \frac{P - P_\infty}{\frac{1}{2}\rho U_{ref}^2}$$

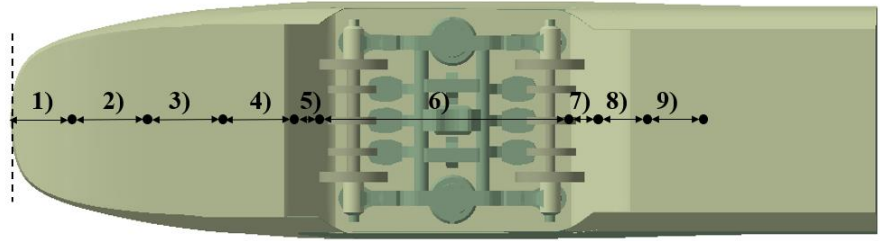

| 1) 0.50m | 3) 0.70m | 5) 0.16m | 7) 0.16m | 9) 0.43m |
| 2) 0.70m | 4) 0.79m | 6) 1.96m | 8) 0.43m | |

**Figure 8** Locations of pressure measurement points

where $U$ is the velocity at that point, $U_{ref}$ the reference velocity, $P$ the pressure at that point, $P_\infty$ the reference pressure, and $\rho$ the density of the flow. The flowfield measurement was obtained without spraying snow. The measurement cases are shown in Table 3. The sampling frequency was 10 Hz for each case.

**Table 3** The cases of flowfield and the surface pressure measurements

| Test case | Free stream velocity | Running time |
|---|---|---|
| Case 1 | 15m/s | 300s |
| Case 2 | 20m/s | 300s |
| Case 3 | 30m/s | 300s |

### 3.2 Snow flux

The spraying of snow flux is a crucial factor that determines the range of snow accretion and the thickness of snow. Therefore, knowledge of the snow flux is required for the experiment. In previous studies, Font (1998) suggested a measuring method in which snow is collected using snow collectors, and this was found to be useful for short-duration measurements. Additionally, Vigano (2012) verified the mass flux of snow for tests using snow collectors. Generally, snow collectors are made of a stainless cylinder and porous tissue that air can pass through, while the snow remains inside the collector. In this study, the diameter and depth of the snow collector were 35 mm and 300 mm, respectively. Nine snow collectors were placed at 4.0 m from the snow nozzle using vertical and horizontal rods and the difference in snow mass was measured before and after the experiment to calculate snow flux. The measurement points were chosen so that the frontal area of the model was covered (Figure 10). Snow flux $\phi_{snow}$ is expressed as follows:

$$\phi_{snow} = \frac{m_{col}}{A_{col} \cdot t}$$

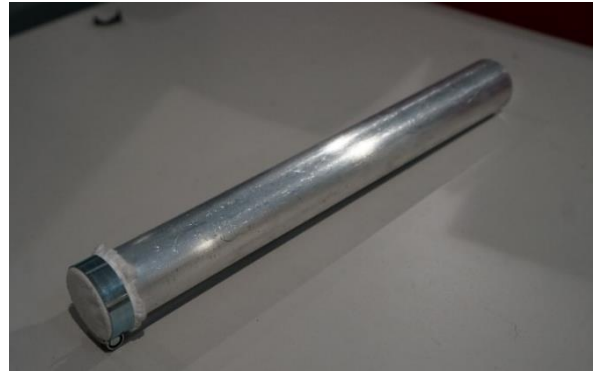

**Figure 9** Image of the of snow collector

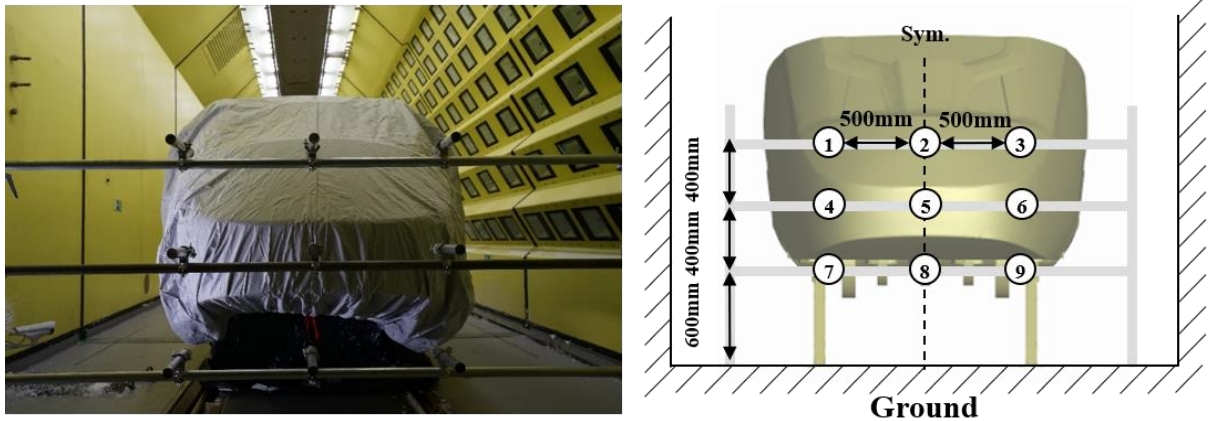

**Figure 10** Arrangement of snow collector

Where $m_{col}$ is the mass of snow accumulated inside the snow collector, $A_{col}$ is the inlet area of snow collector, $t$ the running time. The test conditions for measuring snow flux were a wind speed of 30 m/s and a duration time of 5 min.

### 3.3 Liquid water content of snow

There are a lot of properties to investigate the snow and these are essential to understand the phenomenon of snow accretion. Among the properties of snow, liquid water content (LWC) is defined as the percentage ratio of the water weight to the total weight of snow (Yamamoto (2020)). When there is a high LWC of snow, i.e., wet snow, that has accreted on the structure, the adhesive force due to the hydrogen bonds between the snow and structure is significant. Subsequently, due to bonding between the snow particles, the snow layer grows larger. However, it has been found that dry snow does not stick to vertical surfaces of a model

(Makkonen (1989))). Therefore, the LWC represents an important metric for evaluation of snow accretion.

The key point for obtaining LWC is that the duration of whole process should be completed within a couple of minutes. Because several kinds of heat exchanges occur, the process to collect snow for measuring LWC was conducted in a short period of time and the temperature in the CWT was kept at approximately the setting value, in order to minimize the effect of heat exchanges. However, as described in Section 2.3, the setting speed during the experiment was 30 m/s, in which it is perilous for a person to acquire a snow sample in the wind tunnel. Hence, the wind speed was set to 5 m/s and the collecting location was fixed at 4.0 m from the spraying nozzle to acquire the snow sample. For the purpose of calculating the LWC of the snow sample, the calorimetric method (Kawashima (1998)) was used. The process of measuring LWC was performed in the following sequence.

1. Measure the mass of the empty portable calorimeter and that filled with water. Additionally, measure the temperature of the water in the calorimeter. The water temperature should be between 30 and 40 ºC.
2. Spray snow to an insulated panel. An insulated panel was placed 4.0 m from the spraying nozzle.
3. When the mass snow sample exceeds 100 g using a rough measurement, scrape the attached snow sample on the insulated panel using a scraper.
4. Put the snow in the portable calorimeter and measure the mass and temperature of the mixture of water and snow.
5. Step 3 and 4 should be done within a minute. Finally, calculate the LWC using the portable calorimeter.

All procedures were conducted when the ambient temperature was set to -3, -5, -7, -9 ºC. The trend of LWC with respect to temperature change was confirmed.

**3.4 Snow accretion**

The experimental data have considerable significance in investigating the reason for snow and ice accretion. The data mentioned in the previous section such as flowfield, snow flux, and LWC of the snow sample represent parameters that quantify the experimental environment, allowing for exploration of reasons for the occurrence of snow accretion. In this section, the methods of obtaining the results after the end of the experiment are introduced. Two kinds of method are presented: one is the density of the ice agglomerate and the other is thickness and range of the snow layer.

### 3.4.1 Density of accreted snow

Literature findings have suggested that accretion starts with a thin snow agglomerate, and the growing snow agglomerate is exposed to different kinds of heat exchange, such as convection, evaporation, or condensation of water vapor (Admirat (2008)). As the type of heat exchange typically depends on the ambient temperature, the properties of snow agglomerate depend on the temperature.

In order to acquire the density of accreted snow, one of the characteristics of snow agglomerate, the test piece was attached in front of the left support (Figure 11), as the process of measuring the volume should not affect the snow accretion on the bogie. The test piece had the dimension of 20 mm(width) × 20 mm(height).

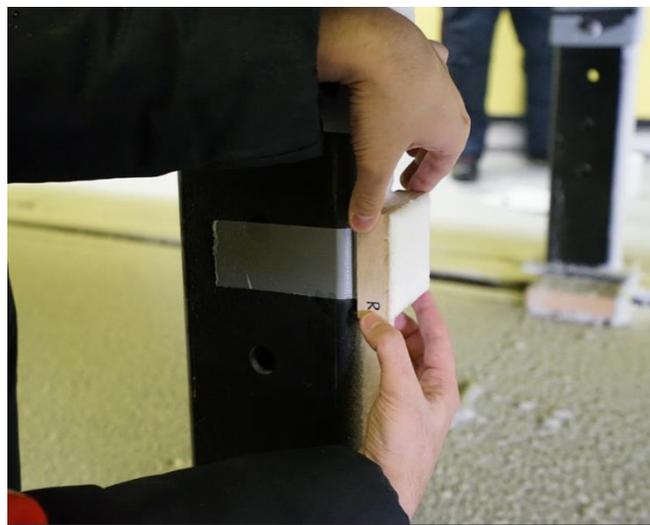

**Figure 11** The test piece for measuring the density of snow sample

Literature findings suggested that the accretion starts with a thin snow agglomerate and growing snow agglomerate is exposed to different kind of heat exchange, such as convection, evaporation or condensation of water vapor (Admirat (2008)). As a kind of heat exchange typically depends on the ambient temperature, the properties of snow agglomerate depend on the temperature. After the snow was accumulated on the test piece, the height of five points on each face was measured using a transparent ruler. Corner nodes were included for each of the five points, and 16 points were measured in total. The volume of accreted snow was calculated based on the measured height. Subsequently, the mass of accreted snow was measured. When measuring mass of the sample, the scale of resolution had an accuracy of 0.1 g. Density was then calculated.

### 3.4.2 Thickness of accreted snow

In the CWT test, documenting the resulting snow accretion is a primary goal. Hence, both quantitative and qualitative data should be included in the results.

There are a number of methods for obtaining quantitative data in the CWT test. Traditionally, the most commonly used method is pencil tracings (Lee (2012); Kozomara (2021)) which requires melting a section of the snow layer and using a board coated with water repellent to trace around the shape. However, this method is used for obtaining only the cross section of the snow layer, meaning that it is not appropriate for obtaining data of the entire area. Therefore, it can be hard to investigate the trend of snow accretion. Additionally, the pencil tracing method involves a risk of damage to the snow layer. When melting a snow layer, it can fall off the surface of model. Hence, in this study, snow accretion on the model was documented using 3D scanning. 3D scans were conducted using a scan arm system, which is certified and has an accuracy of 0.048 mm. Detailed information regarding the 3D scan can be found in a previous work (Kozomara (2020)). Regions for quantitative analysis were the stagnation point at the nose (I), the slope of the nose, and the right side of the bogie (III), as shown in Figure 13

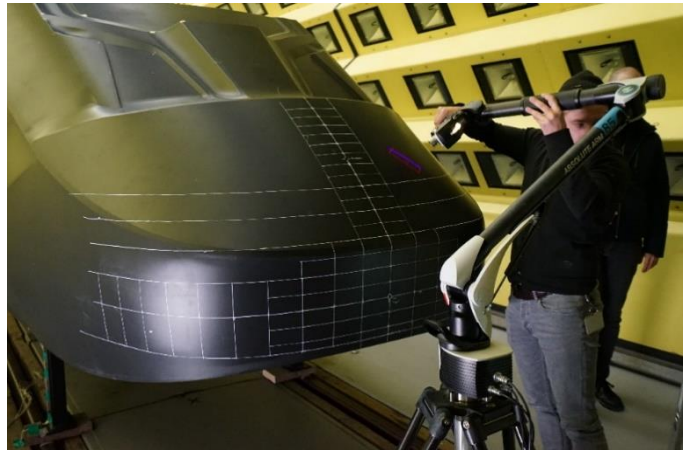

**Figure 12** Scanning process

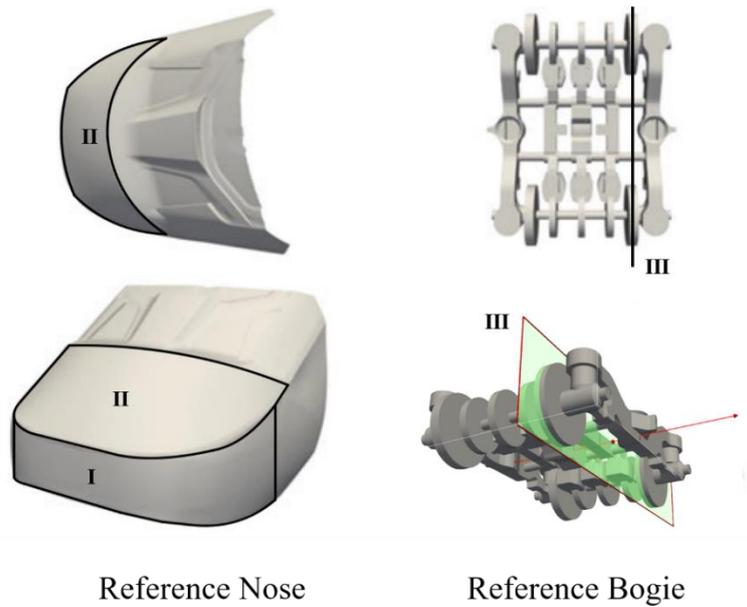

Figure 13 Scanned model for reference and analysis regions

Regarding the qualitative data, two kinds of camera were used. One was for recording the whole process in the CWT that may occur during experiments, and the other one was for photographing the snow accretion after the experiment had finished. A camera for recording was mounted on a wall that did not affect the experimental results. Furthermore, a camera for photographing was set to 17 locations, so that the snow accretion on the front of model and the bogie could be captured. The locations are shown in Figure 14.

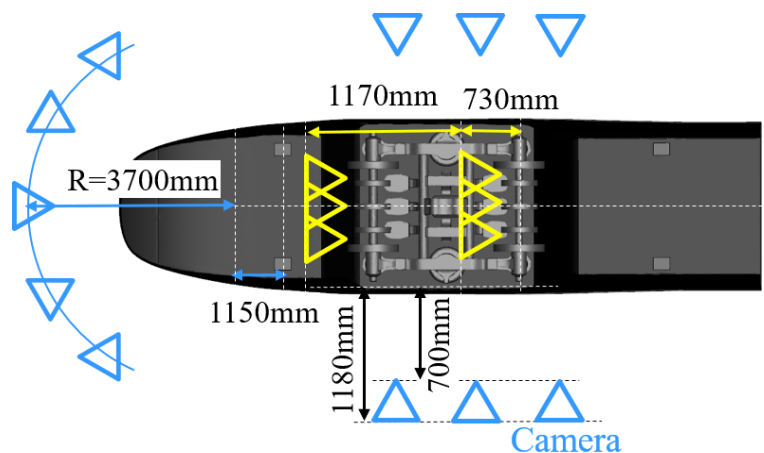

Figure 14 Locations at which photographs were taken

## 4  Experimental result
### 4.1 Flowfield

The velocity distribution on the bottom of the bogie at each location is presented in Figure 15 The velocity ratio mentioned in Figure 15 is expressed as the local velocity divided by the freestream velocity.

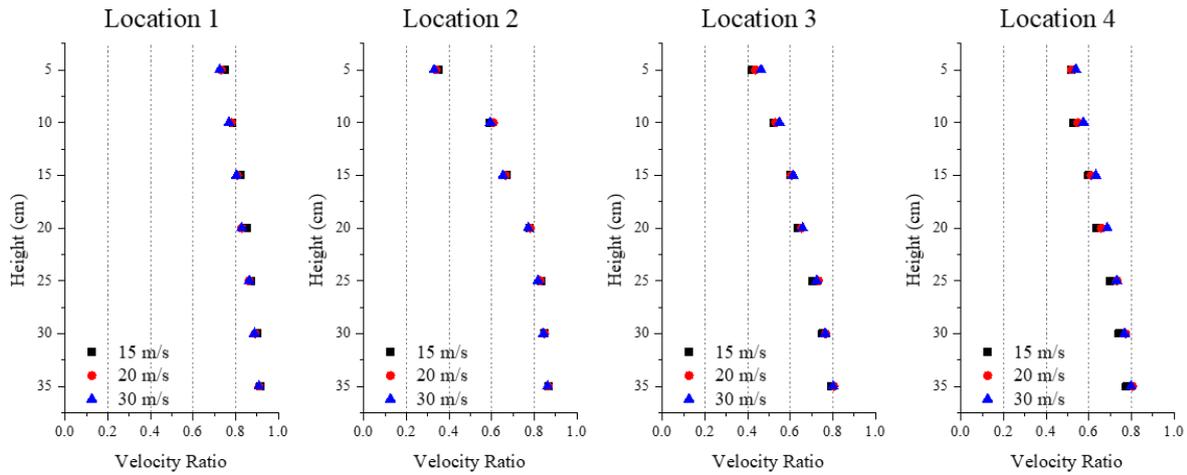

**Figure 15** Velocity distribution at the bottom of the bogie

First, it was found that the velocity ratio was different at each location. The maximum velocity ratio at the point closest to the wall was at location 1. For quantitative analysis, the difference in the velocity ratios between the point closest to the wall (5 cm) and the point farthest from the wall (35 cm) is tabulated in Table 4

**Table 4** Difference in velocity ratios between the nearest and farthest point from the surface of model

| Free stream velocity | Loc. 1 | Loc. 2 | Loc. 3 | Loc. 4 |
| --- | --- | --- | --- | --- |
| 15 m/s | 0.169 | 0.517 | 0.371 | 0.258 |
| 20 m/s | 0.178 | 0.527 | 0.369 | 0.287 |
| 30 m/s | 0.184 | 0.533 | 0.338 | 0.259 |

The difference in velocity ratios between the nearest and farthest point from the surface of model was smallest at location 1 and largest at location 2. At location 1, the geometric shape of model surface is similar to horizontal flat plate and the boundary layer grows normally from the nose. Considering that the range of Reynolds number is $2.1 \times 10^6$ to $4.2 \times 10^6$ and the distance from the nose is 1.94 m, the thickness of the turbulent flat plate boundary layer at location 1 was between 3.5 cm and 4.0 cm. Additionally, the 5 cm point was defined as the point at which the growth of the boundary layer was already completed. Therefore, the velocity ratio at location 1 shows a smooth gradient. However, the other locations have complex geometric shapes, meaning the characteristics of flow are influenced by turbulence and vortices. Especially, the flowfield around location 2 was greatly affected by turbulence and vortices induced by front wheel. Moreover, because the boundary layer thickness is proportional to the distance from the nose, the 5 cm point at locations 2, 3, and 4 can be inside the boundary layer. Furthermore, locations 3 and 4 are on the back of the cavity region and the flow velocity at the cavity is generally small compared to the freestream velocity

(Wang (2018))). Therefore, the gradient of the velocity ratio is larger at locations 2, 3, and 4.

Furthermore, the velocity gradient was demonstrated to be similar according to the freestream velocity. Previous research has suggested that the effect of the Reynolds number is a major factor in determining the flow characteristics (Baker (1991); Schewe (2001)). In this experiment, the range of the Reynolds number was $2.1\times10^6$ to $4.2\times10^6$ As the Reynolds number used in this experiment is within the range required to represent the same phenomenon (Niu (2016)), it was confirmed that there was little difference in the velocity ratio according to the freestream velocity.

Figure 16 presents the pressure coefficient on the surface on each location.

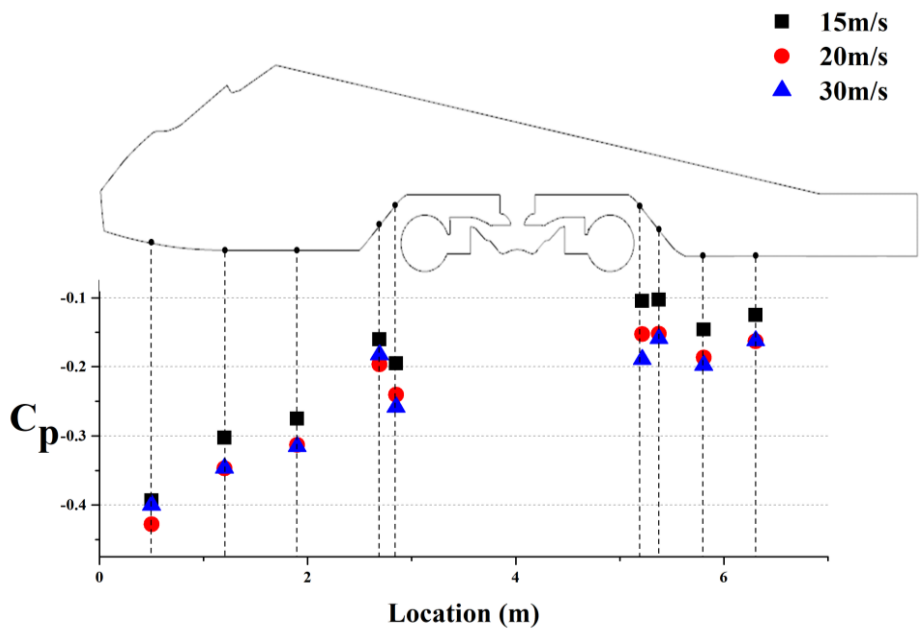

**Figure 16** Surface pressure coefficient at each location

The results demonstrated that $C_p$ was negative at all locations, and it was confirmed that the pressure at the bottom of model was lower than the reference value.

Additionally, the pressure coefficient had the smallest value at the nose, and increased along the longitudinal direction of the model.

Furthermore, there was a stagnation point of the wheel around location 4, and location 5 was within the region in which the flow expands past the stagnation point of the wheel. At the stagnation point, the pressure is a maximum, which decreases as it expands. Therefore, it is considered that $C_p$ at location 5, which is shown in Figure 8, was smaller than that at location 4. Although there is a deviation of the pressure coefficients with respect to the speed, it was found that $C_p$ was similar for each location.

**4.2 Snow flux**

The snow flux distribution at the front of model is shown in Table 5. It was confirmed that the snow flux used in this experiment was concentrated on the right side and the center, which proportions were 34.38 % on the right side and 48.43 % on the center. The averaged snow flux was 24.63 g/m$^2$/s with a standard deviation of 14.54 g/m$^2$/s. Using this data, snow accretion analysis was conducted when considering each case. This will be described in Section 4.5

Table 5 Snow distribution in the front part of the model

| Snow flux (g/m$^2$/s) | Left | Center | Right | Proportion (%) |
|---|---|---|---|---|
| Upper | 13.86 | 48.50 | 17.32 | 35.93 |
| Middle | 13.86 | 24.25 | 13.86 | 23.44 |
| Lower | 10.39 | 34.65 | 45.05 | 40.63 |
| Proportion (%) | 17.19 | 48.43 | 34.38 | |
| Mean snow flux | 24.63 g/m$^2$/s | Standard deviation | | 14.54 g/m$^2$/s |

### 4.3. Liquid water content of snow

The LWC of snow with respect to the ambient temperature is shown as Table 6.

Table 6 LWC for different ambient temperatures at 5 m/s

| Temperature (°C) | -9 | -7 | -5 | -3 |
|---|---|---|---|---|
| LWC (%) | 0 | 0.36 | 8.58 | 34.17 |

The LWC represents a value close to 0 % at -7 and -9 °C, 10 % at -5 °C, and 35 % at -3 °C. This demonstrates that the LWC increased according to the increase in ambient temperature. In the process of spraying water using a nozzle, heat exchanges between the droplet and air occurred through convection, melting of accreted snow, evaporation, and radiation effect. Among them, exchanges by convection and evaporation play the most dominant role in the phase change. In terms of convection, heat exchange was proportional to the difference between the temperature of the air and water droplets in both the cylinder and non-circular cylinders (Admirat (2008); Sparrow (2004)). In this case, the temperature of the water droplets was 0 °C, as they were subcooled, and heat loss from water droplets occurred more actively for a lower ambient temperature. Therefore, phase changes from water to ice occurred and the LWC decreased.

In relation to evaporation, the saturated vapor pressure decreased when the ambient temperature decreases (Lowe (1974)). Moreover, heat exchanges by evaporation was in proportion to the difference between the saturated vapor pressure in air and the vapor pressure of water (Admirat(2008)). Assuming the vapor pressure of water was a constant, the amount of heat exchange by evaporation was larger, and the LWC of water droplets was lower.

Furthermore, as it has previously been found that the LWC decreases when the exposure time increases, it can be suggested that the LWC was higher than approximately 35 % when the flow velocity was larger than 5 m/s, because of the reduced exposure time.

### 4.4 Snow accretion
#### 4.4.1 Density of accreted snow

Table 7 presents a comparison of mass, volume, and density of the snow samples with respect to different ambient temperatures at the same duration time.

| Temperature (°C) | Mass (kg) | Volume (m$^3$) | Density (kg/m$^3$) |
| --- | --- | --- | --- |
| -3 | 7.08×10$^{-2}$ | 8.37×10$^{-5}$ | 845.87 |
| -5 | 4.96×10$^{-2}$ | 8.35×10$^{-5}$ | 594.01 |
| -7 | 2.58×10$^{-2}$ | 4.61×10$^{-5}$ | 559.65 |
| -9 | 1.32×10$^{-2}$ | 2.95×10$^{-5}$ | 447.45 |

**Table 7** Mass, volume, density of snow samples accreted on the front left support

It can be seen that the mass of the snow sample was larger when the ambient temperature was increased. Moreover, the volume and density of the snow sample increased according to the increase in temperature. This can be associated with the LWC discussed in Section 4.3. Yamamoto (2020) suggested that the snow density is proportional to the inverse of $(1 - \Lambda)$, where $\Lambda$ represents the LWC. This demonstrates that the density is closely related to the LWC. Therefore, it can be inferred that the density increases with respect to an increasing LWC. Furthermore, the maximum value of density was 845.87 kg/m$^3$ at -3 °C. which can be related to the density of water. Patterson (1994) calculated the density of water at 1 °C using polynomials based on the combination of interferometric and mechanical measurements, whereby the density was calculated to be 999.90 kg/m$^3$. This shows that the density of the snow sample is close to that of water. Therefore, it can be concluded that the snow sample at -3 °C has less void space, in that water has already filled most of the void space of the snow sample. Therefore, the density increases with respect to the increase in ambient temperature.

#### 4.4.2 Thickness of accreted snow

Results of snow accretion on the nose are shown in Figures 17 to 19. Figure 17 shows a photograph of snow accretion on the nose, Figure 18 is a 3D scan of the snow accretion, and Figure 19 shows the thickness of the snow at the symmetrical plane of the nose.

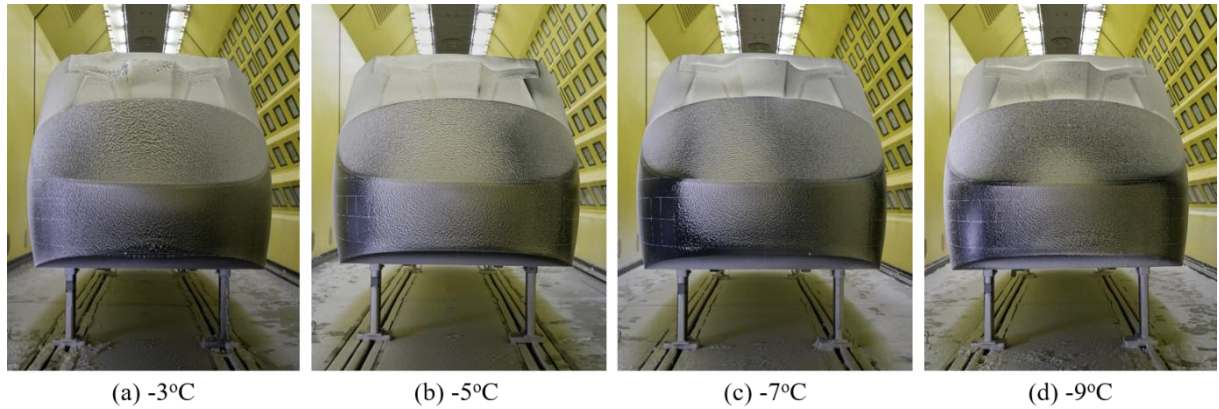

(a) -3°C  (b) -5°C  (c) -7°C  (d) -9°C

**Figure 17** Photographs of snow accretion at the nose

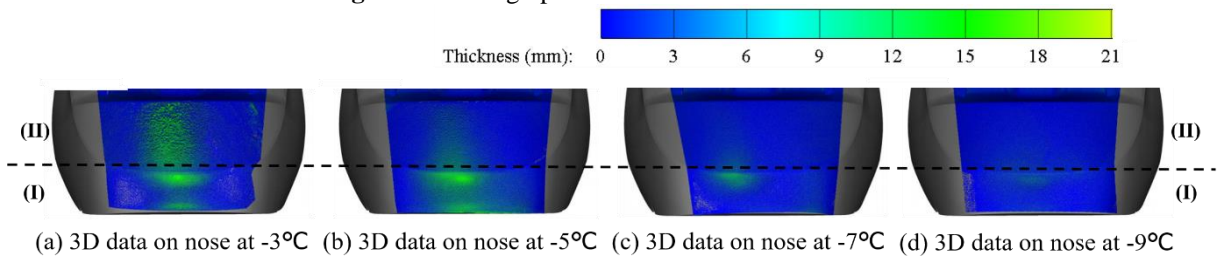

(a) 3D data on nose at -3°C  (b) 3D data on nose at -5°C  (c) 3D data on nose at -7°C  (d) 3D data on nose at -9°C

**Figure 18** 3D scan data of the nose

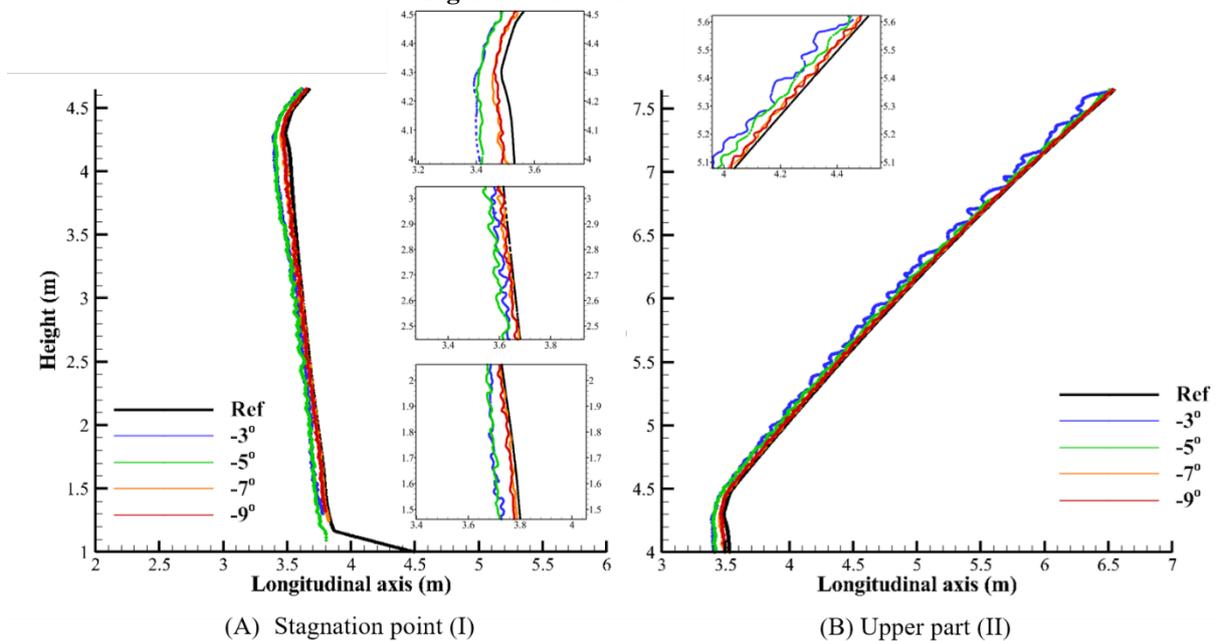

(A) Stagnation point (I)  (B) Upper part (II)

**Figure 19** The thickness of the snow layer at the nose from scanned data

Considering the stagnation point at the nose (I), the grid is barely visible in Figure 17 (a), but becomes clearer as the ambient temperature decreases. This indicates that the snow accretion range is broader with an increasing ambient temperature, and snow more easily accretes when the temperature is high, with subzero temperature. Furthermore, as mentioned in Section 4.3, the LWC is firmly related to the ambient temperature, and it can be suggested that snow accretion is more likely to occur as the LWC increases. This has additionally been presented in previous research (Sakamoto (2000)). For quantitative analysis, the thickness of the snow acquired using a 3D scan, which is shown in Figure 18. It can be seen that the thickness of the snow has a maximum value near the symmetrical plane of the nose. Therefore, the thickness at the symmetrical plane is presented in Figure 19.

Analysis of snow accretion was performed at region (I) according to height, where a height of 4.0-4.5 m represented the upper part of region (I), a height of 2.5-3.0 m represented the middle part, and a height of 1.5-2.0 m represented the lower part. The maximum thickness of each part of the nose and their positions are presented in Table 8.

Table 8 Maximum thickness and location of each part of region (I)

| Temperature (°C) | Upper (4.0-4.5 m) | Center (2.5-3.0 m) | Lower (1.5-2.0 m) |
| --- | --- | --- | --- |
| -3 | 12.51 mm (4.07 m) | 5.22 mm (2.71 m) | 9.27 mm (1.57 m) |
| -5 | 11.61 mm (4.08 m) | 7.15 mm (2.80 m) | 8.80 mm (1.53 m) |
| -7 | 5.15 mm (4.06 m) | 2.30 mm (2.88 m) | 1.50 mm (1.50 m) |
| -9 | 4.31 mm (4.07 m) | 1.58 mm (2.78 m) | 1.57 mm (1.57 m) |

When considering the upper part, the maximum thickness was largest at 12.51 mm at -3 °C, and was located at a height off 4.07 m. It can be observed that the maximum thickness decreases as the ambient temperature decreases. Likewise, the maximum thickness within lower part was largest at 9.27 mm, also at -3 °C, and was located at a height of 1.57 m. This is similar to the snow accretion tendency within the upper part. However, the maximum thickness within the middle part followed a different trend. The maximum thickness was largest at -5 °C, and was 7.15 mm located at a height of 2.80 m. Summing the above results, the cross section of attached snow was trench-like, in that the thickness was smaller in the middle (core) part and was larger in the upper and lower (peripheral) parts, a phenomenon which is similar to previous studies (Koss (2012)). The snow particles collided with the center of the stagnation point, whereby some particles were attached and others were not. Furthermore, particles that did not attach after collision had a lower momentum along the streamline direction around the nose. Finally, particles reattached to the upper and lower part of nose and the icing process was accelerated as time went on. Conversely, the trends of snow accretion at -7 and -9 °C were different to those at -3 and -5 °C. There was no significant difference in thickness between -7 and -9 °C. It was estimated that the LWC at -7 and -9 °C had a very small value, which is one of characteristics of dry snow. Therefore, it can be suggested that the mechanism of snow accretion at -7 and -9 °C is similar.

When considering the slope of the nose (II), snow accretion was observed on the surface and the thickness increased as the ambient temperature increased. Notably, the icing process did not occur symmetrically, and was supposedly due to nonuniform snow flux. Furthermore, the thicknesses at -3 and -5 °C were larger when compared to the thicknesses at -7 and -9 °C.

The flow velocity of region (II) was lower than the reference velocity because of the influence of the stagnation point at region (I). Furthermore, the momentum of particles decreased when moving along the streamline direction. And, these particles were more likely to be attached to the slope than to the stagnation point. Therefore, the thickness of the snow layer at region (II) was larger than that at region (I) in the case of -3 and -5 °C. However, there was no significant difference in the thickness of the snow layer at -7 and -9 °C with respect to region. It can be suggested that the snow accretion in region (II) was nearly related to the LWC, as mentioned in the analysis of region (I).

Figure 20 shows the frontal part of region (III), and Figure 21 shows the rear part of region (III). The left of each subfigure shows the side of region (III).

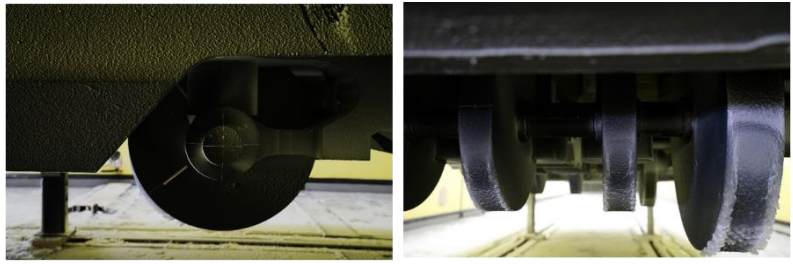

(a) The frontal area of bogie at -3°C

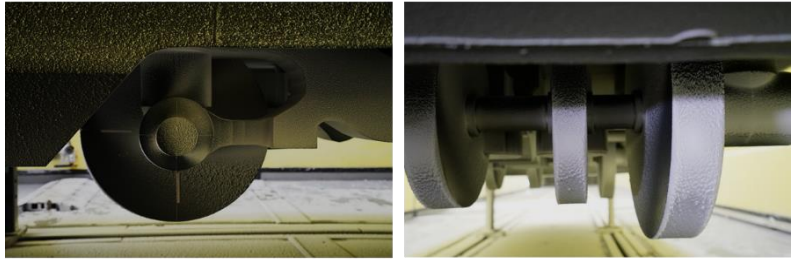

(b) The frontal area of bogie at -5°C

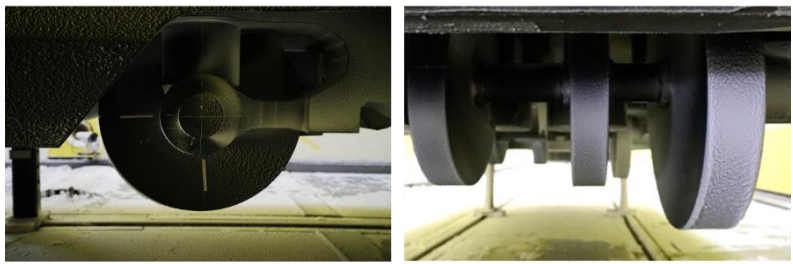

(c) The frontal area of bogie at -7°C

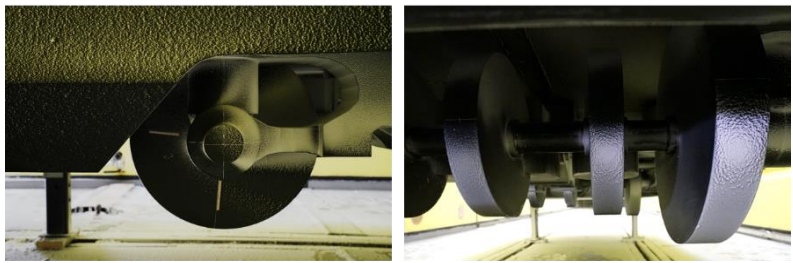

(d) The frontal area of bogie at -9°C

**Figure 20** Photographs of snow accretion on the front of the bogie

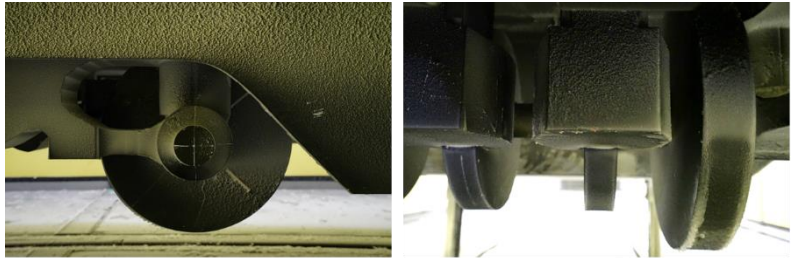

(a) The rear area of bogie at -3°C

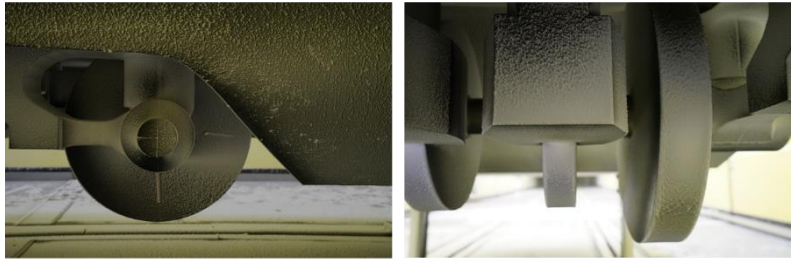

(b) The rear area of bogie at -5°C

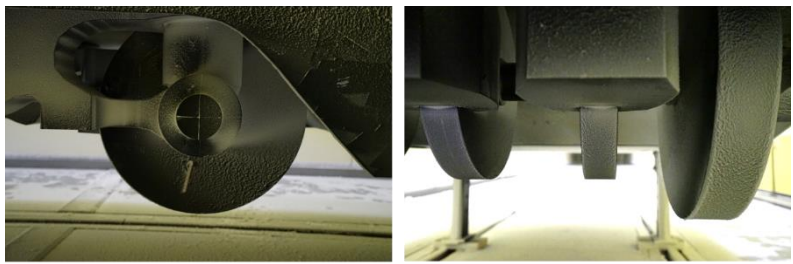

(c) The rear area of bogie at -7°C

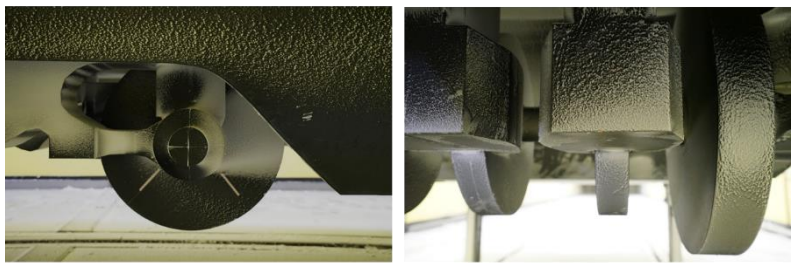

(d) The rear area of bogie at -9°C

**Figure 21** Photographs of snow accretion on the back of the bogie

The snow accreted on all components, including the wheel, brake disc, and bogie frame, as shown in Figure 20, and the break caliper, as shown in Figure 21. It can be seen that the snow accreted on the wheel and brake disc in Figure 20, and 21 (a) is relatively transparent unlike in Figure 20 and 21 (b), (c), and (c). Considering a flow velocity of 30 m/s and the transparency of the snow layer, the LWC should have a value larger than 35 % and the density of snow may be larger than 846 g/m$^3$ at -3 °C, the value presented in the previous Section.

Figure 22 shows the 3D scan data at region (III), and Figure 23 shows the thickness of the snow on the wheel relative to the arc angle.

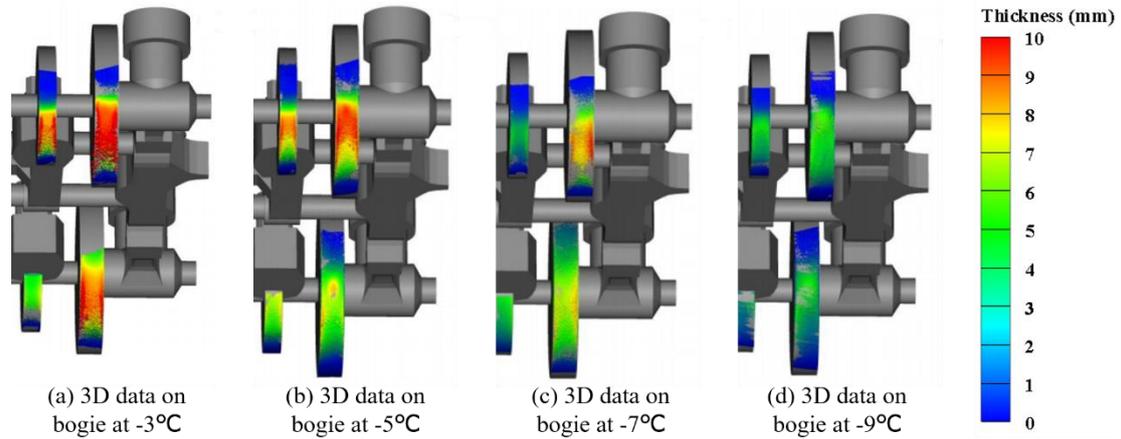

(a) 3D data on bogie at -3°C
(b) 3D data on bogie at -5°C
(c) 3D data on bogie at -7°C
(d) 3D data on bogie at -9°C

**Figure 22** The 3D scan data on the right side of the bogie

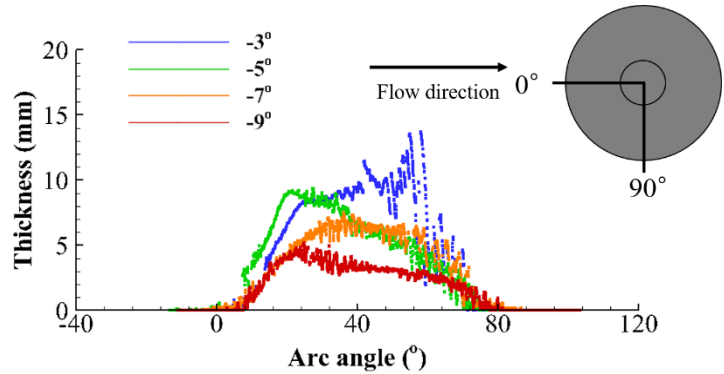

(a) Thickness of snow on front wheel

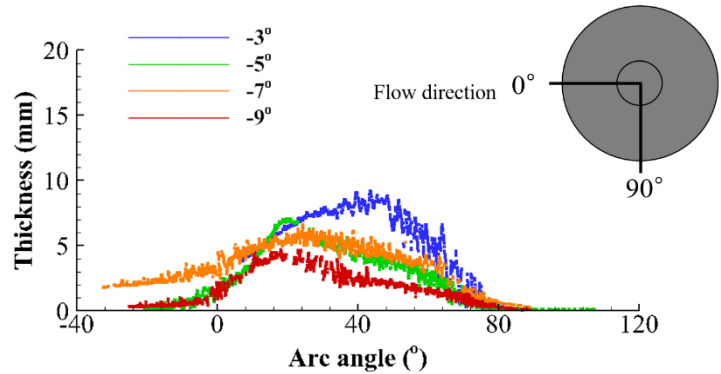

(b) Thickness of snow on rear wheel

**Figure 23** The thickness of the snow layer on the wheel

There are some points to investigate from these results. First, the thickness of snow is a maximum at the edge of wheel. This seems to be analogous with the icing process at region (I). Considering the arc angle is a width component, and given the width and height components, a trench-like structure can be seen, in which the centerline of wheel appears as a hole and the edge appears as the wall. Further, the trench-shaped icing process was demonstrated in all experimental cases. Therefore, the icing process is accelerated at the edge of wheel. Second, the arc angle between 40 to 70 ° showed a maximum thickness on the front and rear wheel at -5, -7, and -9 °C, as shown in Figure 23 (a), and (b). Particularly, a rough icing surface appeared at -3 °C, as shown in Figure 23 (a). Lastly, the thickness of snow on the front wheel was larger than that on the rear wheel at -3 and -5 °C. This is because most of snow particles were attached to the front wheel and the rest of the snow particles were swept downstream and attached to the rear wheel. At -7 and -9 °C, because of the low LWC, the icing process rarely occurred on the front and rear wheel. Therefore, there was no significant difference in thickness.

## 4. Conclusion

In this study, snow accretion according to the ambient temperature was investigated. Using a model of the EMU-320 train scaled by 2/3, a snow accretion experiment was conducted in the climate wind tunnel in Rail Tec Arsenal. Before the snow accretion experiment, data were acquired to investigate properties of snow, and the following conclusions are obtained based on the whole experimental data.

1. The velocity distribution on the bottom of the bogie is different with each location. This is because of the geometric shape of model surface. At location 1, it is similar to horizontal flat plate and the boundary layer grows normally. However, the other locations have complex geometric shapes which can induce turbulence and vortices. Additionally, the velocity distribution was demonstrated to be similar according to the freestream velocity. This is because the Reynolds number used in this experiment is within the range representing the same phenomenon

2. The MVD spectrum of snow particles showed that 90 % of snow particles had a diameter smaller than 77.7 μm. And the most common particle size was a diameter between 40 and 50 μm. Moreover, Snow flux within the experiment was concentrated on the center and the right side of the test section. Additionally, the LWC at a flow velocity of 5 m/s increased with respect to an increase in the ambient temperature. By considering these data, the process of snow accretion was investigated.

3. The density of accreted snow increased with an increase in ambient temperature. The maximum value of density was 846 kg/m$^3$ at -3 °C. Considering that the density at -3 °C was close to the density of water, it is suggested that the snow density and LWC have a similar correlation according to the ambient temperature.

4. Three locations were investigated for snow accretion. At the stagnation point, the snow accretion range was broader with increasing ambient temperature. Further, the cross section of the snow accreted was trench-like, in that thickness at the center of the stagnation point was smaller than that at the upper and lower stagnation points. Considering snow particles on the slope part of the model, it can be suggested that the momentum of particles decreased because of influence of stagnation point. And, snow particles were more likely to be accumulated to the slope than at the stagnation when the ambient temperature increased. At the bogie, snow was accreted on all components including the wheel, brake disc, brake caliper and bogie frame. Considering the snow on the bogie part of the model, trench-shaped icing process was observed on the surface, which was similar to the results at the stagnation point. Moreover, the thickness of the snow on the wheel was a maximum within an arc angle of 40 to 70 º for all cases.

**Declaration of competing interest**

The authors declare that they have no known competing financial interests or personal relationships that could have appeared to influence the work reported in this paper.

**Acknowledgement**

This research was supported by "Development of anti-icing design to prevent snow accretion for high-speed railway vehicle" through the Korea Agency for Infrastructure Technology Advancement (KAIA) funded by the Ministry of Land, Infrastructure and Transport (MOLIT). (22RTRP-B146019-05)

**CRediT authorship contribution statement**

**Wan Gu Ji**: Conceptualization, Methodology, Investigation, Data curation, Writing - original draft, Writing – review editing, Experimental Design. **Soonho Shon**: Experimentation, Methodology. Experimental Design. **Song Hyun Seo**: Experimentation, Methodology, Investigation. **Beomsu Kim**: Experimentation, Data curation. **Kyuhong Kim**: Supervision.

**Declaration of interests**

☒The authors declare that they have no known competing financial interests or personal relationships that could have appeared to influence the work reported in this paper.

☐The authors declare the following financial interests/personal relationships which may be considered as potential competing interests: